# DELAYED LOGISTIC AND ROSENZWEIG–MACARTHUR MODELS WITH ALLOMETRIC PARAMETER SETTING ESTIMATE POPULATION CYCLES WELL

## A. JAN HENDRIKS [1] AND CHRISTIAN MULDER [2] *

[1] Department of Environmental Science, Radboud University, P.O.Box 9010, 6500GL, Nijmegen, The Netherlands

[2] National Institute for Public Health and the Environment, P.O.Box 1, 3720BA, Bilthoven, The Netherlands

*Corresponding author: A.J. Hendriks (A.J.Hendriks@science.ru.nl).*

*\* Submitted by: C. Mulder (christian.mulder@rivm.nl).*

**EXTENDED SUMMARY**

*Context.* So far, theoretical explanations for body-size patterns in periodic population dynamics have received little attention. In particular, tuning and testing of allometric models on empirical data and regressions has not been carried out yet. Here, oscillations expected from a one-species (delayed logistic) and a two-species (Rosenzweig-MacArthur) model were compared to cycles observed in laboratory experiments and field surveys for a wide range of invertebrates and vertebrates. The parameters in the equations were linked to body mass, using a consistent set of allometric relationships that was calibrated on 230 regressions. Oscillation period and amplitude predicted by the models were validated with data taken from literature.

*Models.* The one-species model produced cycle times that increase with organism mass m to the power ¼ if the delay was set equal to the size-dependent age at maturity. If the delay was set on 1 year, the delayed logistic model yielded oscillations with a size-independent period of 4.7 years. Cycle times calculated by the two-species model also scaled to species' body mass with a slope of ¼. The intercepts expected from the two-species were generally higher than those for the one-species model and increased with decreasing consumer-resource body-mass $m_i/m_{i-1}$ ratios. Amplitudes turned out to be size-independent according to both models.

*Results.* The collected data showed that cycle times of herbivores scaled to species' body mass with a slope up to ¼ as expected from the models. With exception of aquatic herbi-detritivores, intercepts were observed at the level calculated by the two-species model. Remarkably, oscillation periods were size-independent for predatory invertebrates, fishes, birds and mammals. Average cycles were of 4 to 5 years, similar to those predicted by the one-species model with a size-independent delay of 1 year. The consistent difference between herbivores and the carnivores could be explained by the models from the small parameter space for consumer-resource cycles in generalist predators. As expected, amplitudes recorded in the field did not scale to size. Observed oscillation periods were generally within a factor of about 2 from the values expected from the models. This demonstrates that one set of slopes and intercepts for age and density parameters applicable to a wide range of species allows a reasonable estimate of independently measured cycle times.

## INTRODUCTION

For a long time, ecologists have been fascinated by the remarkable regularity of population density oscillations observed in laboratory and field studies (e.g., Nicholson 1954; Huffaker 1958; Ginzburg and Inchausti 1997; Turchin 2003). Understanding these cycles is important as at least one-third of the time series display periodicity (Kendall et al. 1999). Even more, understanding such patterns may bring light to the complex dynamics of more common non-cyclic populations (Ginzburg and Inchausti 1997).

While attention for allometric relationships has revived since the late nineties (West et al. 1997 and their subsequent papers), links between the periodicity of population dynamics and body size are still scarce. More than two decades ago, several empirical studies showed already that cycle times for mammalian herbivores scale to their size (Calder 1983; Peterson et al. 1984; Krukons and Schaffer 1991). The objective of the present paper is to underpin these empirical relationships theoretically for a wide range of species. It has already been shown that the period of the oscillations in consumer-resource models scales to body size and temperature (Yodzis and Innes 1992; Vasseur and McCann 2005). Here, we extended (and tried to improve) their work.

Firstly, in addition to the consumer-resource models explored so far, we also included the delayed logistic function in our analysis. We took both models into account because one and two species dynamics cause oscillation of different magnitudes (Jones et al. 1988; Murdoch et al. 2002; Turchin 2003).

Secondly, rather than using a single body size regression for each parameter of the models, we applied a set of allometric coefficients based on a meta-analysis of 230 empirical regressions (Hendriks 1999, 2007; Hendriks and Mulder 2008). Covering a large variety of cold-blooded and warm-blooded taxa, the allometric functions in this study were mutually consistent and complied with other independent data. For instance, the intercept and slope chosen for the consumption rate constant were related to the values chosen for, e.g., the age at maturity and the production efficiency, as all are linked to each other mechanistically.

Thirdly, to test the models' correctness and applicability, we compared their output to independent data, including both the oscillation period and amplitude. In two cases, containing a cold-blooded and a warm-blooded species, predictions and observations were analysed further. For an overall view, the theoretical relationships derived were weighed against empirical regressions, covering a large set of taxa and conditions.

Although the differential equations chosen reflect fundamental processes that can induce realistic oscillations, they do not cover all processes responsible for cycles observed in actual cases. Obviously, other models than the ones used here might be more appropriate in some of these cases. However, our objective was to identify size-dependent patterns, although the parameters were certainly not considered to represent a universal truth. Instead, they were intended as one framework to recognize overall patterns and to identify exceptions. In this paper, we focus on average trends and deviations.

## SPECIFICATION OF EQUATIONS AND DERIVATION OF PARAMETERS

### *Model selection and parameter setting*

Short-term cycles are associated with intraspecific delays as described by first-order, one-species models whereas long-term oscillations are attributed to interspecific trophic interactions, simulated by second-order models with two species or more (Jones et al. 1988; Turchin 2003; Murdoch et al. 2002). For both types, an overwhelming number of differential equations has been proposed (see reviews by, e.g., May 1974 and Turchin 2003). For the purpose of the present study, however, we were confined to models that allow correlation of parameters to body size and analytical solution of the differential equations while still covering the basic mechanisms to produce realistic results. These criteria are met by the delayed logistic model [2] and the Rosenzweig-MacArthur model [12], here chosen as the one-species and two-species model, respectively. As their behaviour has extensively been described, we only briefly recapitalize their main features here.

Instead, our description focuses on the relationship between the parameters of these models and species mass m according to allometric regressions of the kind $\gamma \cdot m^{\pm \kappa}$. In a previous study (Hendriks 2007), elementary rate, time, density and area parameters have been calibrated on 230 allometric regressions to arrive at a mutually consistent set of coefficients $\gamma$ for various clades (Appendix: Table 1). In the present analysis, we applied these relationships to the constants in the models and derived similar functions for more complicated parameters. For convenience, all rate $k$ and time $\tau$ constants were related to the rate constant for mean population turnover $k_p$:

$$k_p = q_T \cdot \gamma_p \cdot m^{-\kappa} \qquad [1]$$

where $q_T$ reflects the temperature correction in comparison with the standard of 20° C (Gillooly et al. 2001). For the overall assessments in the present analysis, temperature T was set on 20° and 37° C for cold-blooded and warm-blooded species, respectively, corresponding to a temperature correction $q_T$ of 1 and 3.5. In the following sections, we describe the links between the parameters of the models and the production constant $k_p$.

## THE DELAYED LOGISTIC MODEL

### *Differential equations and equilibrium*

The delayed logistic model has been studied in detail for internal or external time lags in various disciplines (e.g., Tinbergen 1931; Frisch and Holme 1935; Hutchinson 1948). It is usually denoted as (May 1974)

$$\frac{dN}{dt} = r \cdot \left(1 - \frac{N(t-\tau)}{K}\right) \cdot N(t) \quad [2]$$

Initially, density N increases at a maximum rate r but population growth levels off near the carrying capacity K, reflecting the equilibrium state

$$\varepsilon(N) = K \quad [3]$$

The reduction of the increase does not depend on the actual density N(t) but on the density some time ago N(t-$\tau$), potentially causing overshoots. Here, we explored the periodicity of the delayed logistic model by fixing the time lag on a size-independent value of 1 year ($\tau$=365d) and by setting the interval equal to the non-reproductive period ($\tau$=$\tau_m$).

***Parameters***
The maximum rate of increase r can be estimated as the natural logarithm of the potential lifetime fecundity ln($R_0$) divided by the period between successive generations $\tau_g$, i.e. r = ln($R_0$)/$\tau_g$ (Birch 1948). The generation time or age at average reproduction $\tau_g$ can be calculated as the sum of the juvenile period $\tau_m$ and half of the adult period $\tau_a$ so that the intrinsic rate of increase r equals [4]

$$r = \frac{\ln(R_0)}{\tau_g} \approx \frac{\ln(R_0)}{\tau_m + \tau_a/2} = \frac{\ln(R_0)}{0.5/k_p + (1/k_p)/2} = \ln(R_0) \cdot k_p = \ln(R_0) \cdot q_T \cdot \gamma_p \cdot m^{-\kappa}$$

The life expectancy at maturity $\tau_a$ equals the inverse mortality by definition, i.e. $\tau_a$=1/$k_m$=1/$k_p$. The age at maturity $\tau_m$ varies between 0.3…0.8/$k_p$, depending on the size at hatch and at maturity of the species concerned (Gillooly et al. 2002). Lower values are noted for cold-blooded organisms, whereas the upper end applies to warm-blooded animals. For general derivations, we used the level of

$$\tau_m = \frac{0.5}{k_p} = \frac{0.5}{q_T \cdot \gamma_p} \cdot m^{\kappa} \quad [5]$$

In this perspective, the potential lifetime fecundity $R_0$ thus reflects the maximum capacity of populations to increase under favourable conditions. Proxies of the potential lifetime fecundity $R_0$, such as the number of eggs in a batch or ovary scaled to adult size m with exponents varying widely around 0.5 for cold-blooded species and 0 for warm-blooded species (Hendriks and Mulder 2008).

Its natural logarithm ln($R_0$) will probably be only weakly related to size, if at all, but there are no allometric regressions to confirm this. The number of young delivered in a single reproductive batch ranged between 10 to over $10^5$ for heterotherms and between 1 and 10 for homeotherms but the total reproductive mass delivered in a clutch or litter is a constant fraction of adult weight (Hendriks and Mulder 2008). Apparently, species divert their reproductive effort into many small or a few large offspring. Whereas most offspring may survive until adulthood in artificial situations such as in experiments or farming, extremely favourable conditions that allow maturation of large numbers of offspring ($R_0$>$10^5$) are unlikely.

Fortunately, one can obtain more realistic values of the potential lifetime fecundity $R_0$ under field conditions from a comparison of allometric regressions on the rate of increase r and the average production $k_p$. As both scale to species' body mass with an exponent of -$\kappa$, the ratio of their intercepts can be obtained by rewriting [4] to r/$k_p$=ln($R_0$). Since the intercepts for r and $k_p$ differed approximately a factor of 4 for heterotherms and 1.5 for homeotherms, potential lifetime fecundities $R_0$ were on average expected to be about $e^4$=55 and $e^{1.5}$=4.5 respectively (Hendriks 2007). These levels represent the regenerative capacity of species in their low-density phase better than the occasional case-specific and possibility extreme values reported in literature. The behaviour of the models in the present study was therefore investigated at the means of 55 and 4.5. Filling them in into [4], shows that the maximum rates of increase of equally-sized cold-blooded and warm-blooded organisms differed, on average a factor of about ln(55)·1·$\gamma_p$·$m^{-\kappa}$ / ln(5)·3.5·$\gamma_p$·$m^{-\kappa}$ ≈ 1.2 only.

***Oscillation period and amplitude***
The behaviour of the delayed logistic function can be judged from the relationship

$$\frac{\pi}{2} = r \cdot \tau_m \quad [6]$$

If the rate and delay are small (r·$\tau$<$\pi$/2), the density N proceeds to an equilibrium. If their product is large (r·$\tau$>$\pi$/2), limit cycles emerge (Tinbergen 1931; Frisch and Holme 1935; Hutchinson 1948). In case the delay is caused by the size-dependent non-reproductive period ($\tau$=$\tau_m$), filling [4] and [5] into [6] yields

$$\frac{\pi}{2} = \ln(R_0) \cdot k_p \cdot \frac{0.5}{k_p} = 0.5 \cdot \ln(R_0) \Leftrightarrow$$
$$R_0 = e^{\pi} \approx 23 \quad [7]$$

indicating that the boundary is independent of size. For a low fecundity ($R_0<23$), the model is stable. At high fecundities ($R_0>23$), limit cycles occur. Average values for cold-blooded ($R_0=55>23$) and warm-blooded ($R_0=4.5<23$) species suggest that oscillatory behaviour is more likely to be induced by maturation delays in heterotherms compared with homeotherms. Obviously, the actual fecundity $R_0$ may differ from these averages, depending on the conditions and species concerned. In addition, the age at maturity $\tau_m$ of warm-blooded species is likely to be closer to $0.8 \cdot k_p$ instead of the $0.5 \cdot k_p$ used in the overall analysis here, reducing the boundary value from 23 to $e^{\pi \cdot 0.5/0.8} \approx 7$.

The cycle time $\tau_o$ of the delayed logistic model can be approximated as (Clanet and Villermaux 1989)

$$\tau_o = \left(\frac{e^{r \cdot \tau}}{r \cdot \tau} + 1\right) \cdot \tau$$
$$\approx 4.0...5.4 \cdot \tau \quad (\pi/2 < r \cdot \tau < 2.5) \quad [8]$$

indicating that an oscillation period $\tau_o$ of 4–5 days or years is expected for a diurnal or annual delay $\tau$ (dashed horizontal lines in Figures 1 and 2). Theoretical and empirical evidence for this well-known relationship has been developed in various disciplines (Tinbergen 1931; Frisch and Holme 1935; Hutchinson 1948). If the delay $\tau$ is set on the size-dependent non-reproductive period $\tau_m$, filling [4] and [5] into [8] yields

$$\tau_o = \left(\frac{e^{r \cdot \tau_m}}{r \cdot \tau_m} + 1\right) \cdot \tau_m$$
$$= \left(\frac{e^{0.5 \cdot \ln(R_0)}}{0.5 \cdot \ln(R_0)} + 1\right) \cdot \tau_m$$
$$\approx 3.8...4.1...4.7 \cdot \tau_m \quad (R_0 = 4.5...23...55) \quad [9]$$

The period of the oscillation $\tau_o$ is now about 3.8, 4.1 and 4.7 longer than the age at maturity $\tau_m$, for a fecundity $R_0$ of 4.5 (average warm-blooded), 23 (boundary) and 55 (average cold-blooded), respectively. As the age at maturity $\tau_m$ scales to size, so does the cycle time $\tau_o$ (dashed diagonal lines in Figures 1 and 2). The oscillation amplitude max(N)/min(N) (Lin and Kahn 1980, May 1974) can be estimated from

$$\frac{\max(N)}{\min(N)} = \frac{K \cdot e^{\sqrt{\frac{40 \cdot (r \cdot \tau - \pi/2)}{3 \cdot \pi - 2}} \cdot \cos\left(\frac{\pi \cdot 0}{2}\right)}}{K \cdot e^{\sqrt{\frac{40 \cdot (r \cdot \tau - \pi/2)}{3 \cdot \pi - 2}} \cdot \cos\left(\frac{\pi \cdot 2}{2}\right)}}$$
$$= e^{2 \cdot \sqrt{\frac{40 \cdot (r \cdot \tau - \pi/2)}{3 \cdot \pi - 2}}} \quad [10]$$

As before, if the delay equals the size-dependent non-reproductive period ($\tau=\tau_m$), filling in [4] and [5] into [10] yields

$$\frac{\max(N)}{\min(N)} = e^{2 \cdot \sqrt{\frac{40 \cdot (0.5 \cdot \ln(R_0) - \pi/2)}{3 \cdot \pi - 2}}}$$
$$= 1...21 \quad (R_0 = 23...55) \quad [11]$$

The latter [11] implies that the amplitude max(N)/min(N) is independent of size for fecundities $R_0$ in the range of 23 to 55 (dashed horizontal line in Figure 3).

## THE ROSENZWEIG-MACARTHUR MODEL

### Differential equations and equilibrium

The Rosenzweig-MacArthur model describes the dynamics of a resource i-1 and a consumer i at two adjacent trophic levels as (Rosenzweig and MacArthur 1963)

[12]

$$\frac{dN_{i-1}}{dt} = r_{i-1} \cdot \left(1 - \frac{N_{i-1}}{K_{i-1}}\right) \cdot N_{i-1} - \max(k_{n,i}) \cdot \frac{N_{i-1}^\beta}{N_{i-1}^\beta + N_{50,i-1}^\beta} \cdot N_i$$

$$\frac{dN_i}{dt} = \left(p_{an,i} \cdot p_{pa,i} \cdot \max(k_{n,i}) \cdot \frac{N_{i-1}^\beta}{N_{i-1}^\beta + N_{50,i-1}^\beta} - k_{m,i}\right) \cdot N_i$$

The resource i-1 has the same characteristics as the one-species model, but now without delay. The intake of nutrients by plants and food by animals is described as a fraction $N_{i-1}^\beta/(N_{i-1}^\beta + N_{50,i-1}^\beta)$ of a maximum rate $\max(k_{n,i})$. The half-saturation constant $N_{50,i-1}$ represents the resource density $N_{i-1}$ at half of the maximum intake rate. At a slope $\beta$ of 1, influx is a hyperbolic function of $N_{i-1}$, typical for organisms without alternative resources, such as plants, grazers and specialist predators. The interaction is known as the Monod or Holling type II response for intake of nutrients by vascular plants and for consumption of food by animals, respectively (Monod 1942; Holling 1959). At a slope $\beta$ of 2, influx is a sigmoid function of $N_{i-1}$, i.e. the Holling type III response often observed for generalists that consume different types of food items. A fraction $p_{an,i}$ of the intercepted, absorbed or ingested resource is assimilated by trophic level i. A subsequent fraction $p_{pa,i}$ of the assimilation is converted to production $k_{p,i}$, equalling loss by mortality $k_{m,i}$ in the long term. Interactions with trophic level i-2 and i+1 are considered constant, as reflected by the fixed increase rate constant $r_{i-1}$ and death rate $k_{m,i}$.

The equilibria equal (Rosenzweig and MacArthur 1963)

[13]

$$\varepsilon(N_{i-1}) = (c-1)^{-1/\beta} \cdot N_{50,i-1}$$

$$\varepsilon(N_i) = \frac{p_{an,i} \cdot p_{pa,i} \cdot r_{i-1}}{k_m} \cdot (c-1)^{-1/\beta} \cdot N_{50,i-1} \cdot \left(1 - \frac{(c-1)^{-1/\beta} \cdot N_{50,i-1}}{K}\right)$$

indicating that steady states exist for both trophic levels $\varepsilon(N_{i-1}, N_i) > 0$ if $N_{50,i-1}/K_{i-1} < (c-1)^{1/\beta}$. The ecological scope c is conveniently defined as $p_{an,i} \cdot p_{pa,i} \cdot \max(k_n)/k_m$ and indicates the consumer's potency to keep track of the resource.

### *Parameters*

Starting the allometric relationships for the rate constants, the maximum rate of increase of the resource $r_{i-1}$ has already been set by [4]. On average, production $k_p$ is balanced by mortality $k_m$, so we simply assumed

$$k_m = k_p \qquad [14]$$

The derivation of the maximum intake rate $\max(k_n)$ is somewhat more complex. A fraction $p_{an}$ of the average consumption $k_n$ is assimilated and used for production $k_p$ and respiration $k_r$, thus $p_{an} \cdot k_n = k_p + k_r$. Energy is directed to production with an efficiency $p_{pa}$, viz. $k_p = p_{pa} \cdot p_{an} \cdot k_n = p_{pa} \cdot (k_p + k_r)$. The remaining is used for respiration $k_r = (1-p_{pa}) \cdot p_{an} \cdot k_n = (1-p_{pa}) \cdot (k_p + k_r) = (1/p_{pa} - 1) \cdot k_p$. We assumed that an increase of production requires a proportional amplification of metabolism, i.e. the maximum production $\max(k_p)$ corresponds to a respiration of $\max(k_p)/k_p \cdot k_r$. The maximum production $\max(k_p)$ itself can be obtained from the maximum rate of increase, which equals as the birth minus death rate, usually denoted as $r=b-d$. In mass-equivalent terms this corresponds to the difference of the maximum production minus mortality, viz. $r = b - d = \max(k_p) - k_m = \max(k_p) - k_p$. Maximum production was thus calculated as the sum of the rate of increase and the average production rate, viz. $\max(k_p) = r + k_p$. Following these relationships, the maximum intake rate $\max(k_n)$ can now described as a function of average production $k_p$ according to

$$\max(k_n) = \frac{\max(k_p) + \frac{\max(k_p)}{k_p} \cdot k_r}{p_{an}} = \frac{(r+k_m) \cdot \left(1 + \frac{k_r}{k_p}\right)}{p_{an}}$$

$$= \frac{(\ln(R_0) \cdot k_p + k_p) \cdot \left(1 + \frac{(1/p_{pa} - 1) \cdot k_p}{k_p}\right)}{p_{an}} = \frac{\ln(R_0) + 1}{p_{an} \cdot p_{pa}} \cdot k_p$$

[15]

Applying [15] to the ecological scope c yields

[16]

$$c = \frac{p_{an} \cdot p_{pa} \cdot \max(k_n)}{k_m} = \frac{(\ln(R_0)+1) \cdot k_p}{k_p} = \ln(R_0) + 1$$

In addition to the rate and time constants, the characteristics of the two-species model also depend on density parameters and body-mass ratios. A review on allometric regressions showed that consumer-resource mass body ratios $m_i/m_{i-1}$ were largely independent of the species mass, if invertebrates and vertebrates were separated (Hendriks 1999). Intercepts of these correlations indicated that aquatic invertebrates are about $10^4$ larger than the algae they feed on (Gross et al. 1993). The same value was suggested for ungulates (Shurin and Seabloom 2005). Data for terrestrial insects feeding on herbs and trees were lacking. However, indicative values can be obtained by comparing average consumer and resource sizes. Species mass was in the range of $10^{-4...-2}$ kg for herb-dwelling butterflies, of $10^{-5...-3}$ kg for tree-dwelling butterflies and of $10^{-7...-5}$ kg for other insects (Hendriks and Mulder 2008). With typical size of about $10^{-2...-1}$ kg for herbs and $10^{3...4}$ kg for trees, model estimations with consumer-resource body-mass ratios of about $10^{-2...-1}$ kg, $10^{-8...-7}$ and $10^{-5...-4}$ kg, apply to these insect groups, respectively. In a meta-analysis on predator-prey $m_i/m_{i-1}$ ratios, parasitoids and invertebrate carnivores were, on average, 1-10 times the size of the animals they fed on (Brose et al. 2006). However, vertebrate predators were between $10^2$ and $10^4$ times larger than their prey. Model calculations were carried out with consumer-resource body-mass ratios $m_i/m_{i-1}$ from $10^{-8}$ to $10^4$ for cold-blooded species and at $10^4$ for warm-blooded species. For specific groups, geometric averages of their range were used (Appendix: Table 1).

Because body-size correlations for the carrying capacity K as such have not been reported in literature, we used allometric regressions on density N [kg·km$^{-2}$] instead (Table 1). We assumed that the carrying capacity for a resource equalled the biomass density of all species together within the same trophic level. Theoretical and empirical evidence indicated that densities scale to species mass m with slopes $\kappa$ of about ¼ (Hendriks 2007), hence

$$K \approx N = \gamma_N \cdot m^\kappa \qquad [17]$$

The intercept $\gamma_N$ of the allometric density regression depends on trophic level $i$, number of species $n_i$ and regional conditions, such as the availability of sunlight, nutrients and water. For the global analysis that we seek here, generic values can be derived by assuming that mass density depends on the energy flow available and needed to sustain a trophic level.

Half-saturation constants $N_{50}$ [kg·km$^{-2}$] for nutrients and food have been related to species mass, but slopes $\kappa$ varied between -0.28 and 1.22 depending on the species and size interval covered (Stemberger and Gilbert 1985; Wen et al. 1997). Exponents $\kappa$ tended towards zero for regressions that covered large size ranges, as was observed for average density, although taxon-specific slopes varied around the value of ¼. Allometric regressions suggested $N_{50}/K$ ratios in the range of 1 to 1/10. Values used in model studies are between 1/3 and 1/30 for specific species and in the range of 0 to 1 for general explorations (Yodzis and Innes 1992; Turchin and Batzli 2001; Turchin 2003; Vasseur and McCann 2005). In the present study, oscillation periods will be explored at two different levels, covering differences between typical fecundities and average $N_{50}/K$ values. Using [20] (see next subsection), the analytical solution at the bifurcation was calculated for typical fecundities $R_0$ of 55 (cold-blooded herbivores) and 4.5 (warm-blooded herbivores). Filling in these values into [18], yields the corresponding $N_{50}/K$ ratios of 1/1.5 and 1/2.3, respectively. In addition, numerical simulations were carried out away from the boundary with $N_{50}/K$ at 1/5, being a geometric mean of 1 to 1/30, the range observed. For carnivores ($\beta=2$), oscillations were not expected as underpinned in the next subsection.

### Oscillation period and amplitude

The boundary conditions of the Rosenzweig-MacArthur model are (Yodzis and Innes 1992)

$$\frac{N_{50,i-1}}{K_{i-1}} = \frac{c-1}{c+1} \quad (\beta=1)$$
$$\frac{N_{50,i-1}}{K_{i-1}} = \frac{(c-1)^{1/2}}{1-\frac{c}{2}} \quad (\beta=2) \quad [18]$$

Near these equilibria, densities proceed towards a steady state when $\beta=1$ if $N_{50,i-1}/K_{i-1} < (c-1)/(c+1)$ and $\beta=2$ if $N_{50,i-1}/K_{i-1} < (c-1)^{1/2}/(1-c/2)$ (Yodzis and Innes 1992). In the opposite cases, unstable equilibria emerge. Filling in [16] into [18] reduces the boundary criterion to a relationship between the fecundity $R_0$ and the ratio of the half-saturation constant versus the carrying capacity $N_{50}/K$, according to

$$\frac{N_{50,i-1}}{K_{i-1}} = \frac{\ln(R_{0,i})+1-1}{\ln(R_{0,i})+1+1} = \frac{\ln(R_{0,i})}{\ln(R_{0,i})+2} \quad (\beta=1)$$
$$\frac{N_{50,i-1}}{K_{i-1}} = \frac{(\ln(R_{0,i})+1-1)^{1/2}}{1-\frac{\ln(R_{0,i})+1}{2}} = \frac{(\ln(R_{0,i}))^{1/2}}{1-\frac{\ln(R_{0,i})+1}{2}} \quad (\beta=2)$$
[19]

For $\beta=1$, the $N_{50}/K$ needed for stability rapidly proceeds towards 1, indicating that limit cycles occur for virtually all realistic parameter values. For $\beta=2$ and $R_0>3$, the final term $\sqrt{\ln(R_0)}/(1-(\ln(R_0)+1)/2)$ is negative and thus below $N_{50}/K$, implying that the model is always stable under these conditions.

The oscillation period that can be approximated by (Yodzis and Innes 1992)

$$\tau_o = \frac{2\cdot\pi}{\sqrt{\left|\frac{c-1}{c+1}\cdot r_{i-1}\cdot k_{m,i}\right|}} \quad (\beta=1)$$
$$= \frac{2\cdot\pi}{\sqrt{|(c-1)\cdot r_{i-1}\cdot k_{m,i}|}} \quad (\beta=2) \quad [20]$$

For $\beta=1$, filling in [4], [12], [16] into [20] yields

$$\tau_o = \frac{2\cdot\pi}{\sqrt{\left|\frac{\ln(R_{0,i})}{\ln(R_{0,i})+2}\cdot\gamma_p^2\cdot\ln(R_{0,i-1})\cdot q_{T,i-1}\cdot q_{T,i}\cdot m_{i-1}^{-\kappa}\cdot m_i^{-\kappa}\right|}}$$
$$= \frac{2\cdot\pi\cdot m_i^\kappa\cdot\left(\frac{m_i}{m_{i-1}}\right)^{-\kappa/2}}{\gamma_p\cdot\sqrt{\frac{\ln(R_{0,i})}{\ln(R_{0,i})+2}\cdot\ln(R_{0,i-1})\cdot q_{T,i-1}\cdot q_{T,i}}}$$
$$\approx 3.5...10\cdot 10^3\cdot m_i^\kappa\cdot\left(\frac{m_i}{m_{i-1}}\right)^{-\kappa/2} \quad (1<q_T<3.5, R_0=4.5, 55)$$
[21]

The period of these oscillations increases with the mass of the consumer, $\tau_o \propto m_i^\kappa$ (dashed diagonals in Figures 1 and 2). Equation [21] also shows that the cycle time decreases with temperature, $\tau_o \propto \sqrt{q_{T,i-1}\cdot q_{T,i}}$ and the consumer-resource body-mass ratio, $\tau_o \propto (m_i/m_{i-1})^{-\kappa/2}$ both lowering the intercepts in the graphs. The intercepts calculated by the model differ a factor of $3.5...10\cdot 10^3\cdot(m_i/m_{i-1})^{-\kappa/2}$ due to differences in metabolism ($q_T=1, 3.5$) and fecundity ($R_0=55, 4.5$) of consumers and resources. For the same combination of cold-blooded or warm-blooded species, the expected intercepts differ a factor $(m_i/m_{i-1})^{\kappa/2} = 10^{4\cdot 1/4/2} = 3.1$ for a consumer-resource body-mass ratio $m_i/m_{i-1}$ of $10^4$. For warm-blooded species, the estimations for the bifurcation ($N_{50}/K=1/2.3$) as well as for the average ratio between the half-saturation constant and the carrying capacity ($N_{50}/K=1/5$) were plotted. For cold-blooded species, only levels expected at the bifurcation ($N_{50}/K=1/2.3$) and the standard temperature of 20°C were graphed.

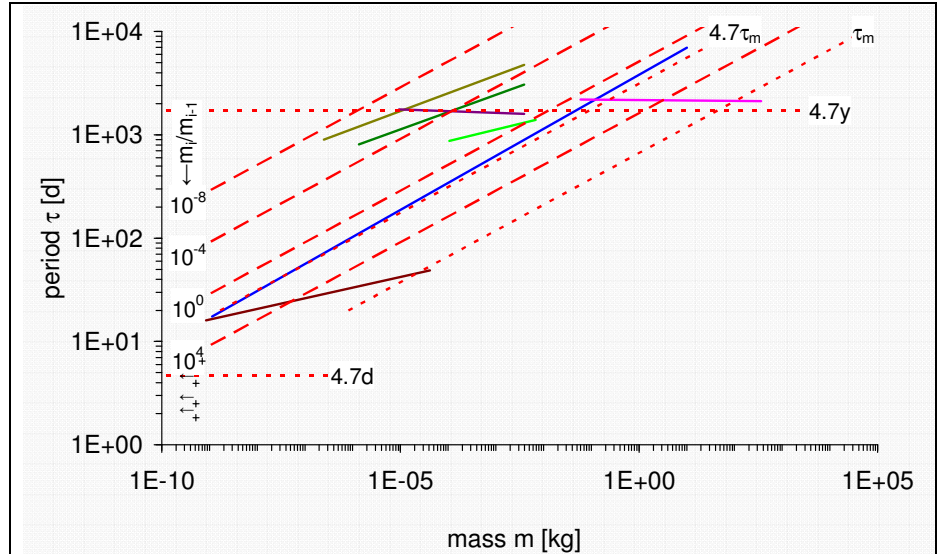

**Figure 1.** Age at maturity $\tau_m$ and oscillation $\tau_o$ period versus species mass $m$ for cold-blooded species. Analytical approximation from the delayed logistic model (red dotted lines - - -) with size-independent (labelled 4.7d and 4.7y, [8]) and size-dependent (labelled $\tau_m$ and 4.7$\tau_m$, [9]) time lag. Analytical approximation from the size-dependent Rosenzweig-MacArthur model with consumer-resource body-mass ratios $m_i/m_{i-1}$ of $10^{-8}$, $10^{-4}$, $10^0$ and $10^4$ at T=20° C (dashed lines, [21]). Regressions (solid lines) obtained from a previous study (Hendriks and Mulder 2008) for species with food added continuously or iteratively (dark —), aquatic herbi-detritivorous invertebrates (blue —), butterflies living on herbs (light green —), butterflies living on trees (dark green —), phytophagous insects living on crops (light brown —), predatory insects (purple —) and fish (pink —). Individual data for unicellulars with masses in the $10^{-19...-9}$ kg range (plusses +).

**Figure 2.** Maturation $\tau_m$ and oscillation $\tau_o$ period versus species mass $m$ for warm-blooded species. Analytical approximation from the delayed logistic model (dotted red lines - - -) with size-independent (labelled 4.7y, [8]) and size-dependent (labelled $\tau_m$ and 3.8$\tau_m$, [9]) time lag. Analytical and numerical approximations from the size-dependent Rosenzweig-MacArthur model with a consumer-resource body-mass ratios $m_i/m_{i-1}$ of $10^4$ at T=20° C, for the bifurcation $N_{50}/K=1/2.3$ and the average $N_{50}/K=1/5$, respectively. Regressions (solid lines) obtained from a previous study (Hendriks and Mulder 2008) for avian and mammalian herbivores (light green —) and carnivores (pink —).

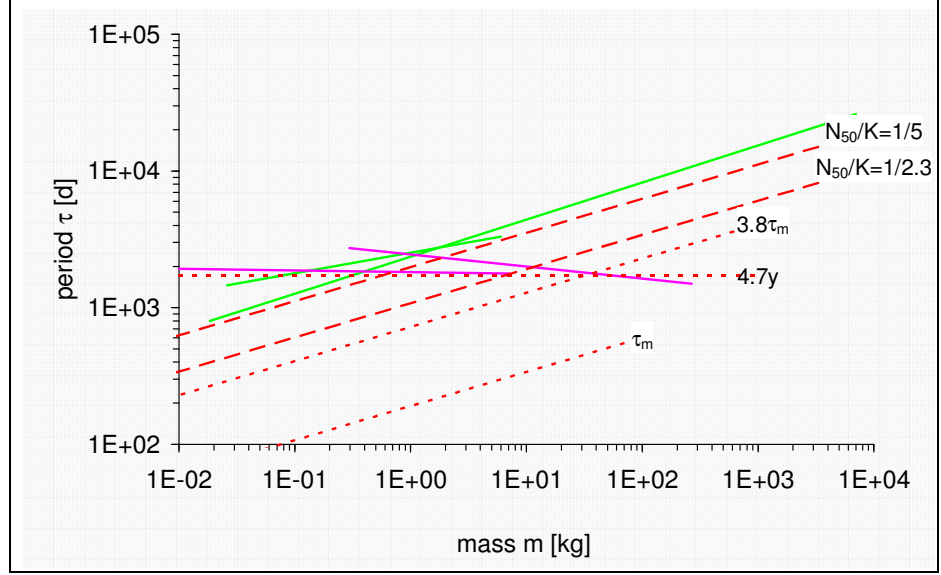

Analytical solutions for the cycle amplitude of the Rosenzweig-MacArthur model are not possible. However, because all rate and density constants scale to size with the same absolute value for the exponent, the ratio of the maximum density, max(N), and the minimum density, min(N), is expected to be constant, which was confirmed by numerical simulations (dashed lines in Figure 3 in box below).

Our allometric parameter setting of the one-species model thus suggested that oscillation periods increase with the mass of the species to a power κ of ¼ (dashed diagonals in Figures 1 and 2). Similar trends were anticipated by our two-species model but for herbivores and specialist carnivores only (Holling type-II, $\beta=1$). Cycles induced by consumer-resource interactions were not expected for generalist carnivores (Holling type-III, $\beta=2$). Size-independent oscillation periods may be attributed to diurnal or annual fluctuations that are not related to species mass (dotted horizontal lines in Figures 1 and 2). Both models suggested that amplitudes are size-independent (dotted and dashed lines in Figure 3 in box below).

## RESULTS

### Validation by a detailed analysis of two species

After having derived and explored the allometric models, we compared the projected dynamics to that observed in lab and field studies, starting with a detailed analysis of two species. To cover different types of species and conditions, we arbitrarily selected a laboratory study on the rotifer *B. rubens* with regular food supply and a field-based investigation on the lemming, *Lemmus sibiricus* (Halbach 1979, 1984; Turchin and Batzli 2001). To facilitate interpretation, we compared our models to observations as well as to similar models that were independently derived by the authors reporting the data.

For *B. rubens*, differences between the delayed logistic model calibrated by Halbach (1984) on laboratory experiments and our independently derived allometric version were small (Table 2). The maximum rate of increase r calculated by us, overestimated the measured value by a factor of 0.82/0.36=2.3. The estimation however, was within the range noted in related experiments. Moreover, at the observed lifetime fecundity $R_0$ of 10, instead of the typical value of 55 for cold-blooded species, the model predicted a maximum rate of increase r of 0.47 $d^{-1}$. The carrying capacity K observed for *B. rubens* in experiments with daily refreshed medium and food was much higher than anticipated by our model and obtained in chemostats with related species (Yoshida et al. 2003). As the carrying capacity determines the level at which the population oscillates but not the period $\tau_o$ and amplitude max(N)/min(N) themselves, we did not adjust the parameters. The observed period $\tau_o$ and amplitude max(N)/min(N) was close to the value expected from our Table 2. The difference between the product r·$\tau_m$ as set in our model and the product observed in the experiments thus had a small influence on the oscillation characteristics.

For *L. sibiricus* and its food, i.e. mosses and small grass, we compared our allometric two-species model to the Rosenzweig-MacArthur version developed by Turchin and Batzli (2001) using observations in Alaska. With the exception of one, all parameter values expected from the allometric relationships were in the ranges derived in that study (Table 3). The outlier applied to the death rate constant $k_m$. In the approach followed here, the death rate $k_m$ was set equal to average production $k_p$, whereas Turchin and Batzli (2001) estimated the mortality rate from the level of consumption at which birth is balanced by death. Their Rosenzweig-MacArthur model produced oscillations with a period $\tau_o$ of 16 years, substantially above the 5 years calculated by detailed models and observed in the field (Turchin and Batzli 2001, our Table 3). The cycles obtained by our allometric version ranged from 3.4 years for the analytical approximation at the bifurcation ($N_{50,i-1}/K_{i-1}$= =1/2.3) to 5.6 years for the numerical solution ($N_{50,i-1}/K_{i-1}$=1/5).

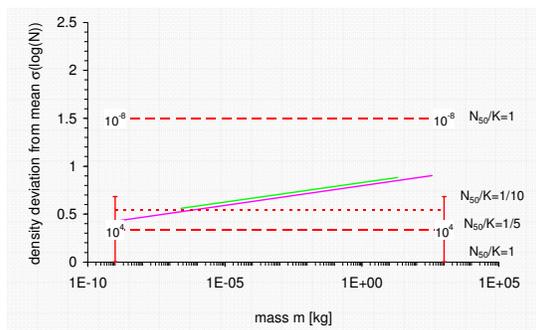

**Figure 3.** Oscillation amplitude σ(log(N)) versus species mass m for cold-blooded and warm-blooded species. Analytical approximation from the delayed logistic model with size-dependent time lag (red dotted lines - - -, [9]). Numerical approximations from the size-dependent Rosenzweig-MacArthur model with a consumer-resource body-mass ratios $m_i/m_{i-1}$ of $10^{-8}$ and $10^4$ at T=20°C with $N_{50}/K$ as indicated. Regressions thereof for herbivores (solid light green line —, σ = 0.82+0.02·log(m), n=87, $r^2$=0.02, p=0.2) and carnivores (solid pink line —, σ = 0.72+0.03·log(m), n=86, $r^2$=0.02, p=0.06). Sources: Kendall et al. (1998) for amplitude, Hendriks and Mulder (2008) for species' body mass.

Results for the amplitude $\max(N_i)/\min(N_i)$ were similar to those for the oscillation period $\tau_o$. Our allometric Rosenzweig-MacArthur model yielded a minimum-maximum density range of 67, which is an order of magnitude below the 400-600 obtained from detailed model studies and field surveys (Table 3). However, setting $N_{50}/K$ to 1/6.9 in our model already produced oscillations with an amplitude $\max(N_i)/\min(N_i)$ of 500, still corresponding to a cycle time of 7 years only.

Both examples demonstrate that the allometric parameter setting for the delayed logistic model and the Rosenzweig-MacArthur model allowed calculation of cycle periods and amplitudes that are within a factor of two from more specialized and detailed models based on field data. Obviously, this type of models does not hold during all phases of the cycles: such analyses would in fact require models that are more comprehensive with case-specific parameter settings. The accuracy margins noted in these two cases do justify a global validation of the generic version on oscillation periods and amplitudes observed in other species and conditions.

### Validation by an overall analysis of oscillation periods

Having explored the credibility of our models in specific cases, the validity of their overall behaviour was also assessed by comparing theoretical predictions to empirical regressions. Below, we subsequently describe the different conditions and taxa covered, discussing similarities and deviations.

Few data on chemostats suggest that cycle times of unicellulars vary around the level of about 4.7 days, expected from diurnal delays in the logistic model (Figure 1). Laboratory experiments in which food was added iteratively had oscillations with a period that increase with size. The scaling slope however, was less than the level of ¼ expected from our delayed logistic model. To identify the cause of this discrepancy, we compared each cycle time measured to the value predicted by our model (Table 4). The fluctuations in laboratory populations of the mite *Eotetranychus sexmaculatus* were about half as frequent as expected, in the absence as well as presence of a predatory mite. For sheep blowflies, *Lucilia cuprina*, fed by ground liver, oscillation periods $\tau_0$ were between 35 and 46 days, caused by egg-adult delays $\tau_m$ claimed to be 9–11 and 13–15 days, respectively (Nicholson 1954; Readshaw and Cuff 1980). The average levels projected by our delayed-logistic model were substantially higher. However, data on age at maturity $\tau_m$ already varied within an order of magnitude around the average calculated from the allometric regression (for insects, see for instance Fenchel 1974).

Egg-adult periods $\tau_m$ of species with a size similar to that of sheep blowflies were therefore in the range of $37 \cdot 10^{\pm \frac{1}{2}}$=12…120. The deviation from our model was thus largely caused by the rapid development of sheep blowflies in comparison to the average noted for equally sized species. Populations of *B. rubens* in daily-refreshed media fluctuated with periods similar to those predicted by our one-species model (Halbach 1979, Halbach et al. 1983, our Table 4). In these experiments, a temperature increase of 10°C reduced the oscillation period by a factor of 2.6 (Halbach et al. 1983), where our model predicted a value of 2.2, expected from the doubling of the metabolic rates ($q_T$ in Table 1). Chemostats with *Chlorella vulgaris* and *Brachionus calyciflorus* also cycled within a range of frequencies that included the estimation by the two-species model. Similar results were obtained for *Daphnia magna*.

Cycle times observed for aquatic herbi-detritivores correlated well to mass (Figure 1). The regression for this group agreed exactly with the prediction of the delayed logistic differential equation for a size-dependent maturation lag. The same trend was also expected from the Rosenzweig-MacArthur model for consumer-resource body-mass ratios $m_i/m_{i-1}$ between $10^0$…$10^4$.

Data for terrestrial herbivorous heterotherms applied to field studies with invertebrates in the $10^{-6}$…$10^{-3}$ kg weight range (Figure 1). Within this small size interval, oscillation periods varied around the 4.7 y level, but scaled to body mass too. Butterflies living on herbs had shorter cycles than equally sized tree-dwelling species, as expected from the Rosenzweig-MacArthur model for realistic herbivore-plant body-mass ratios $m_i/m_{i-1}$ between $10^{-8}$…$10^0$. As size only explained a small part of the variability (9%≤$r^2$≤25%), accurate predictions are unlikely.

For example, populations of the larch bud moth, *Zeiraphera diniana*, fluctuated with cycles of about 9 years (Baltensweiler 1971). The model calculated a period of 13 years at the standard temperature of 20°C. However, at an annual temperature of 7°C, which more realistic for the alpine habitat of this butterfly, [20] yielded a cycle of 29 years.

The slopes noted for frugivore/herbivore birds and mammals were similar to those for cold-blooded herbivores (Figure 2). The intercepts for homeotherms were about twice as high as expected from the analytical solution near the bifurcation ($N_{50}/K=1/2.3$) of the two-species model. However, at the average level ($N_{50}/K=1/5$), numerical simulations converged to the empirical observations. The intercept predicted by the one-species model was 3–4 times lower than noted in the field. The delay imposed by the juvenile period $\tau_m$ of birds and mammals is unlikely to explain the observed patterns as the intercept expected from the one-species model was 3–4 lower than the measured level.

Cycle times for carnivorous heterotherms and homeotherms were not related to size (Figures 1 and 2). Instead, the values varied around the levels predicted for delays of one year, being 4.7 and 3.8, respectively. The invertebrate group was too small to permit subdivision. Splitting up fish into orders did not yield correlations to size (data not shown separately). The oscillation periods for seals, the largest mammalian predators included in the data set, were close to the maturation period $\tau_m$. Due to these two data, cycle time decreased slightly with size for this group.

The empirical regressions thus show that oscillation periods for herbivore animals depended on size, as expected from the one-species or two-species model. Regressions for aquatic herbi-detritivores and warm-blooded herbivores had slopes close to the expected value of ¼. Exponents for invertebrates feeding on herbaceous and woody vegetation were somewhat lower. The intercepts for all herbivores were largely within a factor of two.

Cycle times for carnivores were size-independent and were very close to the level predicted by the one-species model with a size-independent delay of 1 year. In addition, oscillation periods observed for experiments with food was added iteratively or actually interacting with the consumers were within a factor of 2 from the predictions by the one-species and two-species model, respectively.

### *Validation by an overall analysis of the oscillation amplitudes*

The amplitude is usually defined as the ratio between the maximum and minimum density max(N)/min(N) or the standard deviation of the logarithmic density $\sigma(\log(N))$. As the latter is far more frequently measured in field surveys, we carried out numerical simulations with both models and calculated the standard deviation from the densities in the runs. The amplitudes were taken from the same study that provided most of the field oscillation periods (Kendall et al. 1998).

Amplitude expressed as $\sigma(\log(N))$ was hardly correlated to species' body mass m (Figure 3). Size-dependence was also absent in other subdivisions of the data. These empirical results are backed by theoretical evidence. The analytical solution of our one-species differential equation, i.e. [10], and numerical simulations with our two-species model yielded cycles with a size-independent magnitude.

The amplitudes $\sigma(\log(N))$ of herbivores and carnivores were remarkably similar. The observed level could be approximated well by the delayed logistic model with a typical parameter setting for cold-blooded animals (Figure 3). As described-above, the one-species model did not yield limit cycles for a "typical" bird or mammal because of the average fecundity $R_0$ was low. For typical aquatic and ungulate herbivores with a grazer-plant body-mass ratio $m_i/m_{i-1}$ of about $10^4$ and a $N_{50}/K$ range of 1/5 to 1/10, our Rosenzweig-MacArthur model calculated a variability $\sigma(\log(N))$ between 0.3 and 0.7. Amplitudes grew larger with decreasing consumer-resource body-mass ratios $m_i/m_{i-1}$ and $N_{50}/K$.

At a $N_{50}/K$ of 1, for instance, the amplitude $\sigma(\log(N))$ increased from nearly 0 to 1.5 for consumer-resource ratios $m_i/m_{i-1}$ of $10^4$ and $10^{-8}$, respectively (Figure 3). The value of 1.5 indicated that the two-species model could produce realistic minimum-maximum ranges at the order of magnitude level, even for consumers that are substantially smaller than their resource, such as parasites like the eastern spruce budworm, *Choristoneura fumiferana*, and the gypsy moth, *Lymantria dispar* (the highest values in Figure 3). The Rosenzweig-MacArthur model is highly sensitive to changes in $N_{50}/K$; even a small decrease yields to unrealistically large fluctuations.

### **DISCUSSION AND CONCLUSIONS**

Following our objective, we analysed relationships between population density and species size. Parameters of two popular models were linked to body mass. Empirical regressions and individual data on oscillation periods were largely within a factor of two of the levels predicted by our models. Laboratory experiments with food added iteratively yielded cycles in the range predicted by the delayed logistic model for singles species. Field observations on herbivores were at the level expected from the Rosenzweig-MacArthur consumer-resource model, with exception of cycles for aquatic herbi-detritivores that can be understood from the one-species model too. Oscillation periods for carnivores did not scale to size and vary around the level expected from synchronization to annual cycles. The observed amplitudes were related to size for neither herbivores nor carnivores, as expected from the models. Both the one-species and two-species model produced realistic minimum-maximum density ranges but the calculated amplitudes were sensitive to parameter settings.

The slopes and intercepts derived in the present analysis are generally backed by previous studies, where available. The trends predicted and observed in the present analysis correspond to widespread classifications for cycles. Short-term oscillations with a period between 1 to 4 times the maturation period ($1 < \tau_o/\tau_m < 4$) and 1 to 2 times the generation time ($1 < \tau_o/\tau_g < 2$) are generally associated with intraspecific delays and described by first-order, one-species models (Jones et al. 1988, Turchin 2003, Murdoch et al. 2002). The oscillation periods $\tau_o$ calculated by [8] of our delayed-logistic model were also 3.8–4.7 times above the time lag, being one year or the size-dependent juvenile period $\tau_m$. Empirical regressions close to this level were noted for aquatic herbi-detritivores (size-dependent) and all carnivores (size-independent). The data for the other taxa were substantially above the level expected from the delayed logistic model, confirming the absence of size-dependent scaling for warm-blooded expected from [7].

Long-term cycles that take more than 6 times the maturation period ($\tau_o/\tau_m > 6$) and more than 3 times the generation time ($\tau_o/\tau_g > 3$) are usually attributed to interspecific, trophic interactions and simulated by second order, two-species models (Jones et al. 1988; Turchin 2003; Murdoch et al. 2002). Size-dependent oscillations of this magnitude were also predicted for herbivores and specialist carnivores by the allometric Rosenzweig-MacArthur model used here. Empirical regressions for all herbivorous groups were in this range, with exception of the aquatic invertebrates mentioned-above.

The size-dependence of consumer-resource oscillations in the Rosenzweig-MacArthur model has been demonstrated before. Cycle times were increased with the consumer and resource mass according to $\tau_o \propto m_{i-1}^{1/8} \cdot m_i^{1/8}$ (Yodzis and Innes 1992). Rewriting [20] as $\tau_o \propto (m_i/m_{i-1})^{-1/8} \cdot m_i^{1/4}$ provides the same result. The empirical regressions displayed a clear distinction between size-dependent oscillation periods for herbivores and size-independent cycle times for carnivores. As demonstrated before (Yodzis and Innes 1992), our Rosenzweig-MacArthur model suggests that limit cycles for population densities of herbivores and specialist carnivores ($\beta=1$) are possible in a wide range of parameter values [19]. By contrast, generalist predators ($\beta=2$) can only oscillate in very small parameter spaces. Indirect evidence for such a difference comes from latitudinal gradients of periodicity (Turchin and Hanski 1997; Kendall et al. 1998; Klemola et al. 2002). In general, true consumer-resource interactions are likely to be found among specialists in relatively simple ecosystems, such as laboratory or semi-field experiments, agricultural and sylvicultural communities and some biomes, especially tundras.

As noted before, generalists that can be categorized easily and unambiguously, tend to dominate data sets on carnivores (Murdoch et al. 2002). In the present analysis, only two species, *Lynx canadensis* and *Mustela vison*, have been classified with certainty as true specialist carnivores and their oscillation periods were in line with allometric expectations. With the Rosenzweig-MacArthur model indicating that oscillations are unlikely for generalist predators, periodicity observed for this group has to be attributed to other causes, with annual cycles as a first candidate. The delayed-logistic model shows that delays of one year induces a period of about 4 to 5 years, indeed at the average level observed for size-independent oscillations in both cold– and warm-blooded carnivores (Figures 1 and 2).

We did not find reports in literature that refuted or confirmed the absence of a size correlation for the amplitudes found in the present investigation. Indirectly however, the ratio of the maximum and minimum density was reported to be independent of the oscillation periods (Ginzburg and Inchausti 1997). The Rosenzweig-MacArthur model is known to overestimate the amplitude observed in some laboratory and field studies. The discrepancy is attributed to the absence of a compartment for the resource that decreases the amplitude. This includes inedible algae, inaccessible roots or distant areas that are not grazed (McCauley et al. 1999; Turchin and Batzli 2001).

The magnitude of the fluctuations in the Rosenzweig-MacArthur model is very sensitive to the half-saturation constant $N_{50}$. For warm-blooded herbivores and carnivores, we selected values that were close to the measurements and fit well in a general framework explaining differences between trophic levels. The corresponding $N_{50}/K$ ratio of 1/5 produced amplitudes are near the observed level, but the oscillation period was slightly below the empirical regression (Figure 2). Half-saturation constants for cold-blooded species were much more variable, allowing a $N_{50}/K$ interval of 1/1 to 1/10. Within this wide range, realistic cycles can be achieved for consumer-resource body-mass ratios $m_i/m_{i-1}$ between $10^4$ and $10^{-8}$. Whether the model provides a correct explanation can only be established after half-saturation constants have been determined more accurately.

In addition to uncertainties of the models and parameters, the data used for testing are limited. The number of laboratory experiments with, usually aquatic, invertebrates was small, restricting validation for cycles that are shorter than one year. With these model and data uncertainties in mind, the present study showed where of body size dependent oscillations are expected and actually found. Patterns previously observed in data and model studies, largely on herbivorous homeotherms, were generally confirmed for a much wider range of species. Even more, the present analysis showed that cycle characteristics could usually be predicted within a factor of two from the observed values. We consider this surprisingly accurate, keeping in mind that all parameters were set by body size. Finding correct parameter values for a case study of a cyclic population is not easy but demanding rate and time coefficients to be consistent with previous allometric work is even more challenging. We consider our coherent parameter set to be a substantial improvement over the usual practice of picking a single allometric correlation for a specific parameter based on a few related species. Apparently, a single set of values for slopes and intercepts of various rate, age and density parameters applicable to a wide range of species allows a reasonable estimate of independent data on cycle time and, to a lesser extent, amplitude. It indicates that the allometric parameter set might also be more useful for studying the generic behaviour of other models and phenomena than the average, species-specific or arbitrary values that are applied.

In the present analysis, we have focused on average patterns while deviations have been discussed briefly, as it is our belief, rules need to be established before exceptions can be see seen. As such, the current model may serve as a rough check of the population dynamics observed or expected in laboratory experiments and field surveys. Our own future efforts however, will be directed towards quantifying variability and understanding outliers by means of a comprehensive sensitivity and uncertainty analysis. We expect this dynamic model to be also useful for understanding the impact of natural factors and environmental stressors.

# APPENDIX – TABLES 1-4

**Table 1.** Variables and parameters, with typical values for coefficients and exponents derived from allometric regressions as described and explained in text. [Equation numbers in brackets.]

| symbol | description | unit | calculated from/used with typical value of |
|---|---|---|---|
| $\gamma_p$ | average production coefficient | $kg^{\kappa} \cdot d^{-1}$ | 0.00075 |
| $\gamma_N$ | average density coefficient | $kg^{\kappa} \cdot km^{-2}$ | $^p 1.1 \cdot 10^7 \ldots 1.6 \cdot 10^7$, $^c 1.5 \cdot 10^6 \ldots ^w 2.7 \cdot 10^4$ |
| c | ecological scope | - | [16] |
| l | trophic level | - | $^p 1$, $^h 2$, $^k 3$ |
| $k_m$ | mortality (death) rate constant | $d^{-1}$ | [14] |
| $max(k_n)$ | maximum consumption rate constant | $d^{-1}$ | [15] |
| $k_p$ | average production rate constant | $d^{-1}$ | [1] |
| $max(k_p)$ | maximum production (birth) rate constant | $d^{-1}$ | $r + k_m$ |
| $k_r$ | respiration rate constant | $d^{-1}$ | $(1/p_{pa} - 1) \cdot k_p$ |
| K | carrying capacity | $kg \cdot km^{-2}$ | [17] |
| $\kappa$ | allometric exponent | - | ¼ |
| m | mass | kg | variable |
| $m_i/m_{i-1}$ | consumer-resource body-mass ratio | / | $^{tch} 10^{-7.5 \ldots 1.5}$, $^{ach,wh} 10^4$, $^k 10^{1 \ldots 4}$ |
| N | density | # or $kg \cdot km^{-2}$ | [17] |
| $N_{50}$ | half-saturation density | # or $kg \cdot km^{-2}$ | $0.2 \cdot K$ |
| $\varepsilon(N)$ | equilibrium and average density | # or $kg \cdot km^{-2}$ | [3] and [13] |
| $max(N)/min(N)$ | oscillation amplitude | / | [11] |
| $\sigma(\log N)$ | oscillation amplitude | / | |
| $p_{an}$ | fraction assimilated of ingested biomass | $kg \cdot kg^{-1}$ | $^h 0.4$, $^k 0.8$ |
| $p_{pa}$ | fraction produced of assimilated biomass | $kg \cdot kg^{-1}$ | $^c 0.25$, $^w 0.02$ |
| $q_T$ | ratio of rates at T°C and 20°C | / | $e^{\frac{^p 0.3 \ldots ^{cw} 0.6 \cdot 1.6 \cdot 10^{-19}}{1.4 \cdot 10^{-23}} \cdot \frac{T-20}{(T+273)(20+273)}}$ = $^p 0.66 \ldots ^c 0.44\ (10°)$, $^{pc} 1\ (20°)$, $^w 3.5\ (37°)$ |
| $R_0$ | potential lifetime fecundity | $\# \cdot ind^{-1}$ | $^{pc} 55$, $^w 4.5$ |
| r | maximum increase rate | $d^{-1}$ | [4] |
| $\tau_a$ | life expectancy at maturity | d | $1/k_m = 1/k_p$ |
| $\tau_d$ | delay | d | 1, 365 or $\tau_m$ |
| $\tau_g$ | generation time = age at average reproduction | d | $\tau_m + 0.5 \cdot \tau_a = 1/k_p$ |
| $\tau_m$ | age at maturity = age at first reproduction | d | $0.3 \ldots 0.8/k_p \approx 0.5/k_p$ |
| $\tau_o$ | oscillation period | d | [8], [9], [21] |
| T | body temperature | °C | $^{pc} 20$, $^w 37$ |

a = aquatic, c = cold-blooded animals, w = warm-blooded animals, p = producers/plants, h = herbivores, k = carnivores, t = terrestrial. Complete sources mentioned in: Hendriks 1999, Gillooly et al. 2001, Allen et al. 2005, Hendriks 2007.

**Table 2.** Input, parameters and output of the delayed logistic model applied to *Brachionus rubens*. Specific values reported after calibration on empirical data (Halbach 1979, 1984). Generic values apply to the allometric version derived in the present paper.

| factor | | symbol | unit | specific | generic |
|---|---|---|---|---|---|
| input | mass | m | kg | - | $8.5 \cdot 10^{-10}$ |
| | fecundity | $R_0$ | young/adult | 10 | 55 |
| | temperature | T | °C | 25 | 25 |
| parameters | maximum rate of increase | r | $d^{-1}$ | 0.36 [a](0.2-0.7) | 0.82 |
| | carrying capacity | K | $kg \cdot km^{-2}$ | $9.9 \cdot 10^5$ [a]($2 \cdot 10^4$) | $2.6 \cdot 10^4$ |
| | age at maturity | $\tau_m$ | d | 2.0 | 2.4 |
| | age at average reproduction | $\tau_g$ | d | 6.0 | 4.9 |
| output | oscillation period | $\tau_o$ | d | 9.1 | 12 |
| | oscillation amplitude | max(N)/min(N) | - | 10 | 21 |

[a] values from experiments with *Brachionus* (Fernandez-Casalderrey et al. 1991, 1992, Sarma et al. 2001, Yoshida et al. 2003).

**Table 3.** Input, parameters and output of the Rosenzweig-MacArthur model applied to *Lemmus sibiricus* interacting with vegetation at Point Barrow (Alaska). Specific values reported after calibration on empirical data with range for sensitivity analysis in brackets (Turchin and Batzli 2001). Generic values apply to the allometric version derived in the present paper.

| factor | | symbol | unit | specific | generic |
|---|---|---|---|---|---|
| input | resource mass | m | kg | - | [e]$1.0 \cdot 10^{-5}$ |
| | consumer mass | M | kg | - | $1.0 \cdot 10^{-1}$ |
| | resource fecundity | $R_0$ | young/adult | - | 55 |
| | consumer fecundity | $R_0$ | young/adult | - | 4.5 |
| | annual temperature | T | °C | - | -12 |
| parameters | max. rate of increase | $r_{i-1}$ | $d^{-1}$ | $5.5 \cdot 10^{-3}$ ($2.7 \cdot 10^{-3}$-$2.7 \cdot 10^{-2}$) | $1.2 \cdot 10^{-2}$ |
| | carrying capacity | $K_{i-1}$ | $kg \cdot km^{-2}$ | $1.0 \cdot 10^6$ ($2.6 \cdot 10^5$-$2.5 \cdot 10^6$) | $7.5 \cdot 10^5$ |
| | half-saturation constant | $N_{50,i-1}$ | $kg \cdot km^{-2}$ | $3.5 \cdot 10^4$ ($2.5 \cdot 10^4$-$1.0 \cdot 10^5$) | $1.5 \cdot 10^5$ |
| | max. consumption rate constant | $max(k_{n,i})$ | $kg \cdot kg^{-1} \cdot d^{-1}$ | 2.1 (1.4-2.7) | 1.5 |
| | mortality rate | $k_{m,i}$ | $d^{-1}$ | $2.7 \cdot 10^{-2}$ ($1.8 \cdot 10^{-2}$-$3.6 \cdot 10^{-2}$) | $4.7 \cdot 10^{-3}$ |
| | conversion efficiency | $p_{an,i} \cdot p_{pa,i}$ | $kg \cdot kg^{-1}$ | 0.02 | 0.008 |
| output | oscillation period | $\tau_o$ | d | [c]$5.8 \cdot 10^3$ ([d]$2.0 \cdot 10^3$) | [a]$1.3 \cdot 10^3$…[b]$2.0 \cdot 10^3$ |
| | oscillation amplitude | max($N_i$)/min($N_i$) | - | [c]$>8.0 \cdot 10^5$ ([d]$4.0 \cdot 10^2$) | [b]67 |

[a] analytical approximation of [20] ($N_{50}/K=1/2.3$) and [b] numerical simulation of [12] according to the present study ($N_{50}/K=1/5$). Numerical simulations of the [c] Rosenzweig-MacArthur ($N_{50}/K=1/29$) and the more detailed [d] Barrow model by Turchin and Batzli 2001. [e] average of mosses and small grasses.

**Table 4.** Age at maturity $\tau_m$ and oscillation $\tau_o$ period for cold-blooded species in laboratory experiments with food added iteratively or developing interactively with consumer.

| species | mass m [kg] | temp. T [°C] | maturity $\tau_m$ [d] | | oscillation $\tau_0$ [d] added iteratively | | oscillation $\tau_0$ [d] interactively | |
|---|---|---|---|---|---|---|---|---|
| food | | | data | model | data | [a]model | data | [b]model |
| *Eotetranychus sexmaculatus* | $1 \cdot 10^{-9}$ | 28 | 6.0 | 2.2 | 21 | 10 | 71 | [c]35-85 |
| *Lucilia cuprina* | $4 \cdot 10^{-5}$ | 25 | 9.0-15 | 37 | 35-46 | 170 | n.a. | - |
| *Brachionus rubens*, *B. calyciflorus* | $9 \cdot 10^{-10}$ - $1 \cdot 10^{-9}$ | 15 | | | 21 | 26 | | |
| | | 25 | 2.4-2.6 | 2.4-2.6 | 9.5 | 12 | 6.0-26 | 7-18[d] |
| *Daphnia magna*, *D. pulicaria* | $5 \cdot 10^{-7}$ | 16-18 | 10-15 | 21-25 | 100-∞ | 99-120 | 63-180 | [e]49-150 |
| | | 25 | ≈7 | 12 | 51 | 52-57 | | |

[a] one-species model [2], [b] two-species model [12] at $N_{50}/K=1/1.5…1/5$. Sources: Pratt 1943, Nicholson 1954, Huffaker 1958, May 1974, Halbach 1979, Readshaw and Cuff 1980, Halbach et al. 1983, Yoshida et al. 2003, interaction at β=1 with [c] predatory *Typhlodromus occidentalis* ($2.0 \cdot 10^{-6}$ kg), [d] Algae ($m_i/m_{i-1}=10^{-4}$).